\documentclass{elsart}

\usepackage{amssymb,amsmath} 
\usepackage{color}
\usepackage{hyperref}
\usepackage{graphicx}%[dvips]
\newcommand{\beq}[1]{  \begin{equation} \label{#1} }  
\newcommand{\eeq}{     \end{equation}}  

\renewcommand{\appendix}{\setcounter{section}{0}\renewcommand{\thesection}{\Alph{section}}  			\section*{Appendix} 
}

\newcommand{\rf}[1]{(\ref{#1})}
\def\bd#1{\mbox{\boldmath$\displaystyle\mathbf{#1}$} }
\def\dd{\operatorname{d}} 	
\def\tens#1{\mathbb{\,#1}}	
\def\tr{\operatorname{tr}} 										%tensor in 3D
\def\Sym{\operatorname{Sym}}
\def\sgn{\operatorname{sgn}}

\begin{document}%%%%%%%%%%%%%%%%%%%%%%%%%%%%%%%%%%%%%%%%%%%%%%%%%%%
\begin{frontmatter}  %%%%%%%%%%%%%%%%%%%%%%%%%%%%%%%%%%%%%%%%%%%%%%%%%
%\date{September 30, 2006}
%\pagestyle{myheadings}\markright{\sc Singularity at the lid edge}%~~~\today}
%\singlespacing%\doublespacing

\title{Wavefront singularities associated with the conical point in elastic 
	solids with cubic symmetry}% 
\author{Andrew N. Norris}
\address{{Rutgers University, Department of Mechanical and Aerospace Engineering, 98 Brett Road, Piscataway, NJ  08854, norris@rutgers.edu}}
%\maketitle

\begin{abstract}
The wavefronts from a point source in a solid with cubic
symmetry are examined with particular attention paid to the
contribution from the conical points of the slowness surface. 
An asymptotic  solution is developed that is uniform across the 
edge of the cone in real space, the interior of which contains the 
plane lid wavefront analyzed by  Burridge \cite{Burridge67}.  The uniform
solution also contains the regular wavefronts away from the cone edge, 
a delta pulse on one side and its Hilbert transform on the other. 
In the direction of the cone edge the three wavefronts merge to 
produce a singularity of the form $H(t) t^{-3/4}$.  
%The analysis uses the integral representation of Burridge and approximates the slowness surface near the conical point by a curved circular  cone.  

\end{abstract}
\end{frontmatter}  %%%%%%%%%%%%%%%%%%%%%%%%%%%%%%%%%%%%%%%%%%%%%%%%%

\section{ Introduction}

%\cite{Every85,Hurley85,Wolfe98} \cite{Burridge63,Shuvalov00,Buchwald59,Borovikov01,Vavrycuk99} \cite{Paszkiewicz01,Paszkiewicz01b,Paszkiewicz04}

We consider the time dependent Green's function in solids of cubic symmetry, with focus on  the field across the ``flat lid" associated with the conical degeneracy. 
Such conical points  occur in the [111] directions
 in every  material displaying cubic symmetry  \cite{Musgrave}. 
It is well known that the Green's function in anisotropic crystals  can be split into a continuous field and a sequence of
singularities.  The latter are associated with {\em regular}  points on the slowness surface with outward normal in the 
direction of the observer. Burridge 
 \cite{Burridge67} provided explicit expressions for 
the field contributions from regular points, whether they occur
on sections of the slowness surface that are locally convex, concave or saddle
shaped.  He also derived the peculiar singular behavior
on the flat lids of the wave surface associated with the 
conical points.  The  singularity on the flat lid 
is a  step function in time, in contrast to the 
more singular delta pulse or its Hilbert transform (a $1/t$ 
pulse) which arise from regular points on the slowness surface.  
Furthermore, the wavefront on the flat lid decays with distance like $1/r^2$, in contrast to the $1/r$ decay from regular points.  The analysis
in \cite{Burridge67} was limited to receiver positions inside
the lid but not too close to the lid edge.  Here we develop solutions that include Burridge's flat lid solution, but also display the transition across the lid edge.  In particular a uniform solution is derived that  incorporates the regular and the flat lid wavefronts.

%%%%%%%%%%%%%%%%%%%%%%%%%%%%%%%%%%%%%%%%%%%%%%%%%% Figure
\begin{figure}
\begin{minipage}[b]{0.5\linewidth} % A minipage that covers half the page
\centering
\includegraphics[width=9cm]{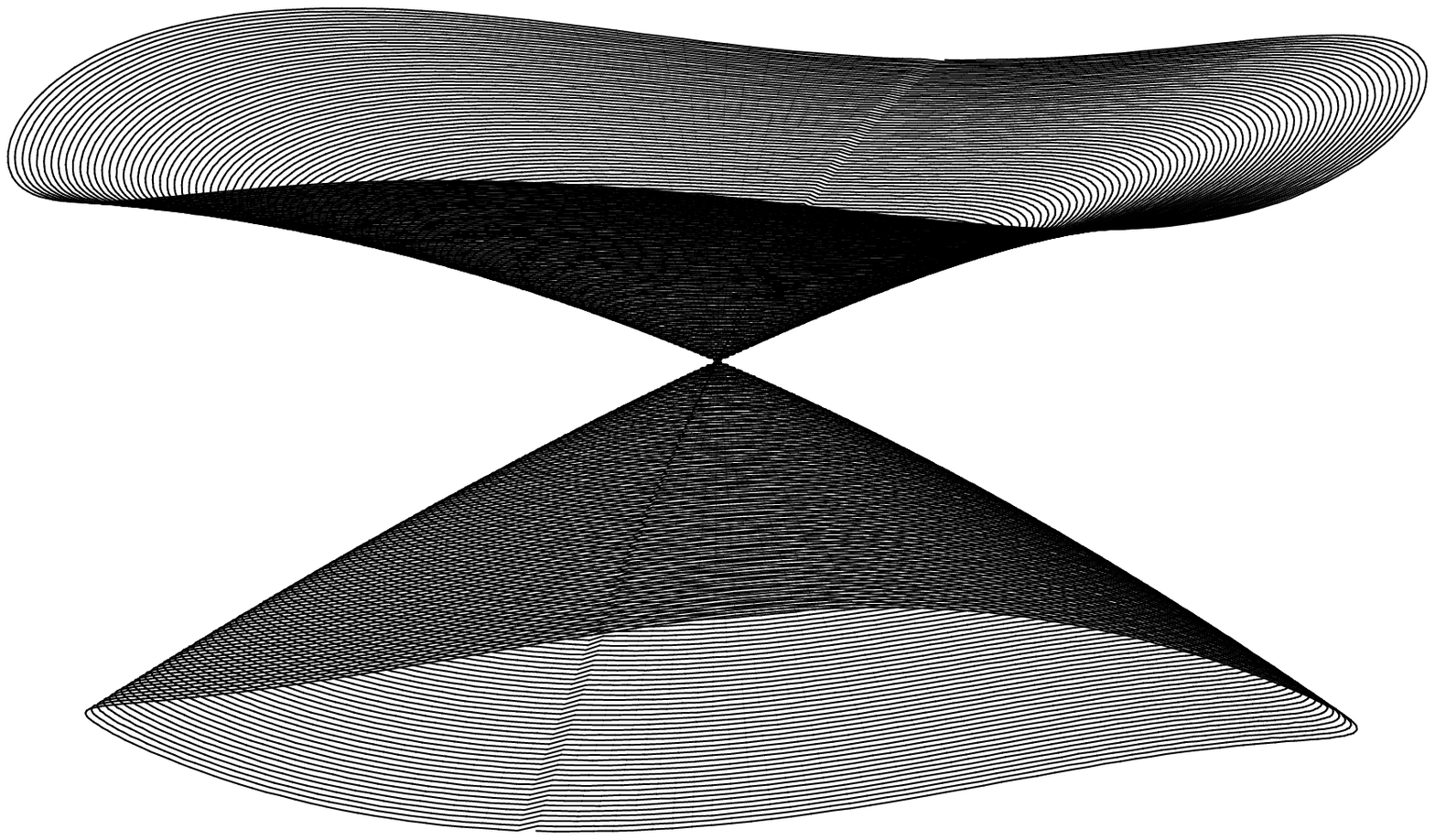}
%\caption{En liten bild}
\end{minipage}
%\hspace{0.5cm} % To get a little bit of space between the figures
\begin{minipage}[b]{0.5\linewidth}
\centering
\includegraphics[width=9cm]{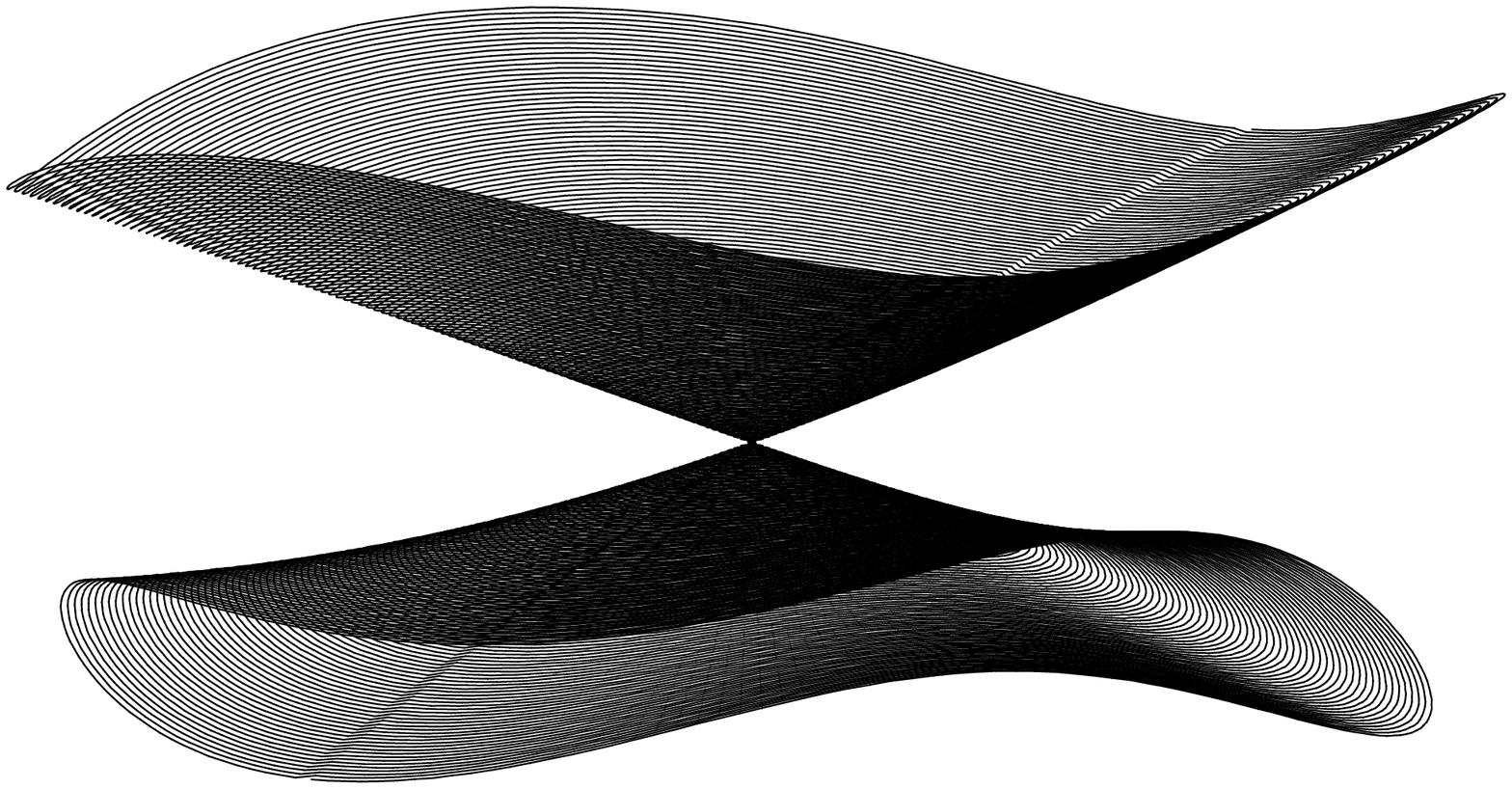}
%\caption{En liten bild till}
\end{minipage} 	
	\caption{The  neighborhood of a conical point on the slowness surface for iron (left) and potassium chloride (right).  The two quasi-transverse sheets are shown for a sector of 7$^\circ$ angular extent about the 111-axis (defined such that $|\xi_\phi/
	\xi_3| < \tan 7^\circ$ in the notation of eq. \rf{0.06}).} 
	\label{fig0}  
\end{figure}
%%%%%%%%%%%%%%%%%%%%%%%%%%%%%%%%%%%%%%%%%%%%%%%%%% 

%The analytic result  \cite[formula (7.22)]{Burridge67} is not valid along the cone which is defined by the edge of the lid (see also equation \rf{1.39} below). 

 %The flat lid singularity does not appear to fall into the general scheme of singularities arising other topological features in the slowness surface, such as kiss singularities \cite{Vavrycuk02}, and it  should not be confused with the quite distinct  phenomenon of internal conical refraction \cite{Burridge06}.  
 Several   studies of the elastic wavefield from conical points appeared after the original  work of Burridge \cite{Burridge67}. Barry and Musgrave 
\cite{Barry79} reduced the solution arising from elliptical conical points  in tetragonal media to a single integral requiring numerical quadrature. Payton 
\cite{Payton92} considered the effects of double points in the slowness surface of a special type of transversely isotropic solid with the  restriction $c_{11}=c_{33}=c_{23}$.    This exhibits conical points on the symmetry axis as well as a ring of double points in the plane perpendicular to the axis.   Based on the material restriction he was able to construct explicit integral solutions for a point force in the form of elliptic integrals which show  that the wave front of conical points have lids, while the circular double points do not produce lids.  Borovikov \cite{Borovikov00b} is the  only work that has considered a uniform solution   valid across the edge of the flat lid.   Borovikov's solution is, however,   time harmonic and not directly applicable to the problem considered here, which we find  more easily addressed by a suitable modification of Burridge's original approach. 

The purpose of this paper is to examine the way in which the 
singularity from the conical point, i.e., the flat lid wavefront, 
coalesces with the nearby regular wavefronts at their common meeting
points along the  directions of the conical boundary in real space.  First we review some
of the necessary preliminary results from 
\cite{Burridge67} and examine the topology of the neighborhood of 
a conical point.  The major results follow from  the use of a locally parabolic approximation near the conical point, which allows certain closed form solutions of the Green's function.  In particular, we derive a uniform approximation that includes the regular wavefronts and the flat lid singularity, and is uniform across the edge of the lid in real space.   The solution also generates the singular form of the Green's function in the direction of the lid edge.  

The outline of the paper is as follows.  The general theory for the Green's function in anisotropic solids is reviewed in section \ref{sec2}. Materials with cubic symmetry are considered in section \ref{sec3} where   the conical point is defined.  The local behavior of the slowness and wave surfaces is discussed in section \ref{sec3}, and  the crucial  slowness surface curvature is introduced.  The  Green's function is reduced to a single integral in section \ref{sec4} and the dominant contributions to the wave field are derived. This includes the flat lid response %\cite{Burridge67} 
and the regular wavefronts associated with the direct waves.  Section \ref{sec5} provides a uniform description appropriate to the conical point, using a generic function which incorporates the flat lid singularity and the regular wavefront singularities.

\section{ General theory for anisotropic solids}\label{sec2}

We are concerned with the fundamental solution, ${\bf G}({\bf x},t)$
with elements $G_{ij}$, which solves 
\begin{subequations}
\begin{align}\label{0.01}
\rho\,\partial_t^2 G_{im}({\bf x},t) - 
C_{ijkl}\, \partial_j\partial_l G_{km}({\bf x},t)  &= \rho\, \delta_{im}
\, \delta(t)\, \delta({\bf x}),
\\
G_{ij}({\bf x},t)&= 0,\quad t<0. 
\end{align}
\end{subequations}
Here $\delta(t)$ is the Dirac delta and $\delta({\bf x})
= \delta(x_1)\delta(x_2)\delta(x_3)$, where the subscripts (the summation convention is understood) 
$i,j, ~etc.$ take on the values 
$1,2$ and $3$ and refer to a rectangular coordinate system.   The elastic
moduli satisfy the symmetries
$C_{ijkl} = C_{jikl}$, $C_{ijkl} = C_{klij}$, and are positive definite, $C_{ijkl}\,a_{ij}a_{kl}>0$
for all symmetric non-zero ${\bf a}$.  Although there are now several alternative ways to express ${\bf G}({\bf x},t)$, e.g. \cite{Every94}, we use the  succinct Herglotz-Petrowski formula \cite{couranthilbert2}  derived by Burridge \cite{Burridge67},  
\begin{subequations}
\begin{align}\label{0.03}
{\bf G}({\bf x},t)  &= \int_S  \dd S\, {\bf F}(\bd{\xi} )\, \delta'(t-\bd{\xi}\cdot{\bf x})
, \\
{\bf F}(\bd{\xi}) &= -\, \frac{1}{8\pi^2}\, 
\frac{ {\bf A}(\bd{\xi}) {\bf A}^T(\bd{\xi}) }{|\nabla v(\bd{\xi})|}, \quad \bd{\xi} \in  S. \label{0.03b}
\end{align}
\end{subequations}
Here, $S$ is the entire slowness surface; $\bd{\xi}$ is the slowness vector on $S$; ${\bf A}(\bd{\xi} )$ is the associated
unit polarization vector; the  phase speed function $v(\bd{\xi} )$ is a homogeneous function of degree one in $\bd{\xi} $, and they are related by 
\beq{0.04}
Q_{ik}({\hat{\bd \xi}}) \, A_k(\bd{\xi}) = \rho\, 
v^2({\hat{\bd \xi}})\, A_i(\bd{\xi}),
\eeq
where
\beq{0.041}
Q_{ik}({\hat{\bd \xi}}) = 
C_{ijkl}\, {\hat \xi}_j {\hat \xi}_l,\qquad 
\mbox{and}\quad \hat {\bd{\xi}} = \bd{\xi}/\xi. 
\eeq
 The eigenvalue equation \rf{0.04}
admits of three roots for $v^2$, corresponding to the three
sheets of the slowness surface. The integral \rf{0.03} is therefore
over the union of the three sheets.  Note that   $v(\bd{\xi})= 1$ on $S$, implying that the physical (dimensional) phase speed is
$v({\hat{\bd \xi}}) = 1/\xi$.   Another property that follows from the functional form of $v$   as a homogeneous function of its argument is that the gradient appearing in 
eq. \rf{0.03b} is the wave (or group) velocity: ${\bf c} = \nabla v(\bd{\xi})$. 

\section{Conical points of the cubic slowness surface}\label{sec3}% for cubic symmetry}

\subsection{Elastic moduli for cubic symmetry}
We use a slightly different notation than that of Burridge \cite{Burridge67}, with the purpose of obtaining a higher order approximation to the slowness surface in the vicinity of the conical point. 
The elastic stiffness is
\beq{85}
 {\tens C} = 3\lambda  {\tens J} + 2\mu_1 {\tens I}+ 2(\mu_2-\mu_1)   {\tens D}, 
\eeq
where the tensors are as follows:  ${\tens I}{\bd S} ={\bd S}$, ${\tens J}{\bd S} = \frac13 (\tr {\bd S} ){\bd I}$ for all ${\bd S}\in \Sym$, and 
${\tens D}$ is a fourth order tensor defined by the cube axes   $\{{\bd a},{\bd b},   {\bd c}\}$, as
\beq{40}
{\tens D} = {\bd a}\otimes{\bd a}\otimes{\bd a}\otimes{\bd a}
 +{\bd b}\otimes{\bd b}\otimes{\bd b}\otimes{\bd b}
 +{\bd c}\otimes{\bd c}\otimes{\bd c}\otimes{\bd c}\, . 
 \eeq
  The stiffness is positive definite if and only if $\mu_1>0$, $\mu_2>0$  and $\lambda + \frac23 \mu_2>0$. 
The non-zero elastic moduli  in the Voigt notation are  $c_{11}=c_{22}=c_{33}=\lambda + 2 \mu_2$, $c_{12}=c_{23}=c_{31}=\lambda$ and $c_{44}=c_{55}=c_{66}=\mu_1$.

\begin{figure}[htbp]
				\begin{center}	
				\includegraphics[width=3.6in , height=3.6in 					]{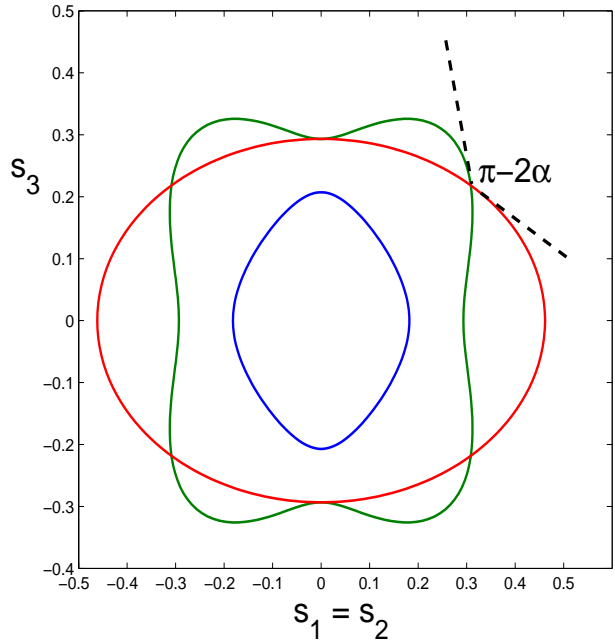} 
	\caption{The  section $s_1=s_2$ of the slowness sheets  for Iron, where ${\bd s}$ is slowness.  The inset depicts the conical section 
	of half angle $\frac{\pi}2 - \alpha$.}
		\label{fig1} \end{center}  
	\end{figure}

The eight conical points on $S$  occur symmetrically
in each octant, although we restrict 
our attention to the conical  point in the $(+++)$ octant. 
Two illustrations of sections of $S$ through the conical point 
can be found in Figures \ref{fig1} and \ref{fig2}.  In order to examine the local behavior of 
$S$ we introduce the new basis set of vectors 
%\begin{align}\label
\beq{0.05}
{\bf e}_1 =\frac{1}{\sqrt{6}}\, (-1,\, -1,\, 2),\qquad
{\bf e}_2 =\frac{1}{\sqrt{2}}\, (1,\, -1,\, 0),\qquad
{\bf e}_3 =\frac{1}{\sqrt{3}}\, (1,\, 1,\, 1). 
\eeq
When referred to these  axes the 
moduli become, using a slight modification of  eq. (3.46) in 
 \cite{AuldI}, 
\beq{0.052}
{\bf c}' = \begin{bmatrix} 
c_{11}' & c_{12}' & c_{13}' & 0 & -c_{46}' & 0 \\
c_{12}' & c_{11}' & c_{13}' & 0 &  c_{46}' & 0 \\
c_{13}' & c_{13}' & c_{33}' & 0 &        0 & 0 \\
 0 &   0 &   0 &  c_{44}' & 0 &  c_{46}'  \\
-c_{46}'  & c_{46}'  & 0&0& c_{44}'  & 0\\
 0 &   0 &   0 &  c_{46}' & 0 &  c_{66}'
\end{bmatrix},
\eeq
where
\begin{align}
c_{11}' & =\lambda+2\mu_2 -(\mu_2 -\mu_1),
& 
c_{33}' & =\lambda+2\mu_2 -\frac43 (\mu_2 -\mu_1),
\nonumber \\
c_{12}' &=\lambda+\frac13 (\mu_2 -\mu_1),
& 
c_{13}' &=\lambda+\frac23 (\mu_2 -\mu_1),
\nonumber \\
c_{44}' &=\mu_2 -\frac13 (\mu_2 -\mu_1),
&
c_{66}' &=\mu_2 -\frac23 (\mu_2 -\mu_1),
\nonumber \\
c_{46}' &=\frac{\sqrt{2}}{3} (\mu_2 -\mu_1) . &&
\nonumber
\end{align}

The acoustical tensor ${\bf Q}  (\bd{\xi})$   defined in  eq. \rf{0.041} 
%$Q_{ik}(\bd{\xi}) = C_{ijkl}\xi_j\xi_l$,    
becomes in the new frame
\begin{align}\label{0.054}
{\bf Q}  (\bd{\xi}) = &
(\lambda+c_{44}' ) \bd{\xi} \otimes \bd{\xi}  + c_{44}' \xi^2 {\bd I} 
- \frac{1}3 (\mu_2 -\mu_1)\, \times \, 
\nonumber \\ & 
\nonumber \\ & \, \, 
 \begin{bmatrix} 
  \xi_1^2 + \xi_2^2 - 2 \sqrt{2} \xi_1\xi_3 &    2 \sqrt{2} \xi_2\xi_3 &   
  \sqrt{2}   (\xi_2^2-\xi_1^2) -2  \xi_1\xi_3 
 \\  && \\
  2 \sqrt{2} \xi_2\xi_3 &    \xi_1^2 + \xi_2^2 + 2 \sqrt{2} \xi_1\xi_3 & ~~
  2 \sqrt{2} \xi_1\xi_2 -2 \xi_2\xi_3  
  \\ && \\ 
  \sqrt{2}   (\xi_2^2-\xi_1^2) -2  \xi_1\xi_3  ~~& ~~2 \sqrt{2} \xi_1\xi_2 -2 \xi_2\xi_3~~   &  2\xi_3^2
%c_{11}' \xi_1^2 +  c_{66}' \xi_2^2 +   c_{44}' \xi_3^2  -2 c_{46}' \xi_3\xi_1 & (c_{12}' +c_{66}' )\xi_1\xi_2 + 2 c_{46}' \xi_2\xi_3 & (c_{13}' +c_{44}' )\xi_3\xi_1 - c_{46}' (\xi_1^2-\xi_2^2)  \\ (c_{12}' +c_{66}' )\xi_1\xi_2 + 2 c_{46}' \xi_2\xi_3 & c_{66}' \xi_1^2 +  c_{11}' \xi_2^2 +   c_{44}' \xi_3^2  +2 c_{46}' \xi_3\xi_1 & (c_{13}' +c_{44}' )\xi_2\xi_3 +2 c_{46}' \xi_1\xi_2 \\ (c_{13}' +c_{44}' )\xi_3\xi_1 - c_{46}' (\xi_1^2-\xi_2^2) & (c_{13}' +c_{44}' )\xi_2\xi_3 +2 c_{46}' \xi_1\xi_2 & c_{44}' (\xi_1^2 + \xi_2^2) +   c_{33}' \xi_3^2 
\end{bmatrix} ,  
\end{align}
where $\bd \xi$ is also referred to the basis vectors of eq. \rf{0.05}. 

%%%%%%%%%%%%%%%%%%%%%%%%%%%%%%%%%%%%%%%%%%%%%%%%%% Figure
%%%%%%%%%%%%%%%%%%%%%%%%%%%%%%%%%%%%%%%%%%%%%%%%%% 

\subsection{The slowness surface near the conical point}

Although formally exact solutions of the Christoffel equation det$ ( {\bf Q} (\hat {\bd \xi}) -\rho {\bf I}) = 0$  are known for cubic crystals \cite{Every79}, we focus here on an approximation that retains the local geometry near the conical point.  The conical point $\bd{\xi}_0= \xi_0{\bf e}_3 $ is a  double root with 
 $\xi_0 = \sqrt{{\rho}/{c_{44}'}}$, or 
\beq{0.045}
\bd{\xi}_0 = \sqrt{\frac{3\rho}{\mu_1+2\mu_2}} \,  {\bf e}_3.
\eeq
The  slowness surface near the conical point 
may be approximated locally as an hourglass-type cone, such that 
\beq{0.06}
\bd{\xi} = \bd{\xi}_0 + 
\zeta {\bf e}_3 + \gamma(\zeta,\phi) {\bf e}_\phi ,
\eeq
where the unit vector ${\bf e}_\phi $ is defined 
\beq{502}
{\bf e}_\phi  = \cos \phi\, {\bf e}_1+\sin \phi\, {\bf e}_2.
\eeq
The coordinates $\zeta$ and $0<\phi\le 2\pi$ will be used to approximate the slowness surface in the vicinity of the conical point $\zeta = 0$.

The approximation used by  Burridge \cite{Burridge67} 
was  $\gamma =  \beta \zeta$, 
corresponding to a right circular cone, 
where $\beta = \cot \alpha$, and $\frac{\pi}{2} - \alpha$ is the semi-vertical angle of the 
right circular tangent cone at the conical point,
\beq{0.065}
\tan \alpha  = \frac{|c_{46}'|}{c_{44}' }
= \sqrt{2} \, \frac{ |\mu_2 - \mu_1|}{\mu_1+2\mu_2}.
\eeq
Note that $0 \le \alpha \le \arctan \sqrt{2} = 55^\circ $, the lower limit  being  the isotropic value, and the upper limit occurs as $\mu_2/\mu_1 \rightarrow 0$. The angle $\frac{\pi}{2} - \alpha$ is  $\nu$ in \cite{Burridge67}. 

%%%%%%%%%%%%%%%%%%%%%%%%%%%%%%%%%%%%%%%%%%%%%%%%%% Figure
\begin{figure}[htbp]
				\begin{center}	
				\includegraphics[width=3.9in , height=3.0in 					]{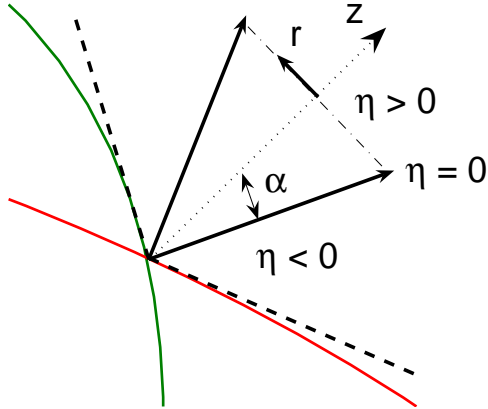} 
	\caption{The conical lid is defined by the cone of semi-angle $\alpha$ in real space.  The coordinate $z$ is length along the [111] direction, and $r$ is the perpendicular distance from the axis.  The parameter $\eta$ is the nondimensional distance from the lid edge, eq. \rf{1.11}. }
		\label{fig2} \end{center}  
	\end{figure}
%%%%%%%%%%%%%%%%%%%%%%%%%%%%%%%%%%%%%%%%%%%%%%%%%% 

The section of the cone in the plane containing the [110] and [001] axes is shown in Figure \ref{fig1}. 
The cone in the slowness sheet is defined by the intersection of the quasi-transverse sheets at a point along the [111] direction.  The associated wave velocity cone in real space is defined by the normals to the cone, which sweep out a cone of semi-interior angle $\alpha$ in real space.  The flat ``lid" of this cone is the circular plane lid under investigation.  We are particularly interested in the form of the wavefront on the lid and across its boundary edge ($\eta = 0$ in Figure \ref{fig2}).  The interior of the lid defines a region in space 
of solid angle $(1 - \cos \alpha) 2 \pi$.  The eight lids of the slowness surface together define a fraction of total solid angle equal to  $8(1 - \cos \alpha) 2 \pi/(4\pi)= 4(1 - \cos \alpha)$, 
which entails as much as 73.4\% of space.  In other words, the region of influence of the conical point resulting from  a point source has influence in, for instance, 37.4\% of space in Iron.  It is important to realize that the field originates from a single point on the slowness surface, and therefore does not generate energy at infinity.  The conical lid wavefront is a nearfield effect. 

In order to  capture the local curvature of the slowness surface 
%evident in Figure ?? 
we assume a parabolic approximation to the curvy cone: 
\beq{1.03} 
 \gamma(\zeta,\phi) =  \beta \zeta   + \frac12 {\kappa (\phi)} \zeta^2.
\eeq
Substituting from \rf{1.03} in 
det$ ( {\bf Q} (\hat{\bd{\xi} }) -\rho {\bf I}) = 0$  and expanding in powers of
$\zeta$, the O($\zeta^2$) term implies \rf{0.065}, while  
the O($\zeta^3$) term yields, in agreement with eq. (28) of \cite{Shuvalov96}, 
\begin{eqnarray}\label{1.035} 
\kappa(\phi) &= &
%\frac{1}{\xi_0 |c_{46}'|} \left\{ c_{66}'-\frac{{c_{46}'}^2}{c_{44}'} +\frac{1}{4}\left[ c_{13}' + c_{33}' - 2\frac{(c_{13}'+c_{44}')^2}{c_{33}' -c_{44}'} \right]\, (1+\cos 3 \phi)\right\} \nonumber \\ &=
\frac{3 (\mu_1+2\mu_2)^2}{2^{3/2} \xi_0  |\mu_2-\mu_1|^3 }
\bigg[ % \left\{
\frac{3\mu_1\mu_2}{\mu_1+2\mu_2}
  \nonumber \\ 
  && \qquad  
-(\mu_2-\mu_1)\big[ 1 +  \frac{8(\mu_2-\mu_1)}{9(\lambda + \mu_1)}\big]
(1-(\sgn(\mu_2-\mu_1)) \cos 3 \phi) \bigg] %\right\}
. 
\end{eqnarray}
Thus $\kappa(\phi)$ displays three-fold symmetry, with extremal values attained
in the directions $\phi=0$ and $\phi=\pi/3$. 
The cone angle and curvature for a range of materials are tabulated
in Table 1.  Note that the curvature for some materials changes signs,
indicating that convex and concave sections may coexist. 
If $\mu_2-\mu_1 >0$ define    $\kappa_0 = \kappa(0)$,  $\kappa_1 =\kappa(\pi/3)$, and if $\mu_2-\mu_1 <0$, reverse the definitions. 
Thus, $\kappa_0>0$, and the curvature changes sign iff the ratio $\kappa_1/\kappa_0$ is negative.  The vanishing of $\kappa$ defines a parabolic line, or a line of zero Gaussian curvature. 
The  examples in Figure \ref{fig0} illustrate the two possibilities: the curvature is always positive for iron but takes both signs for   KCl.   
Parabolic lines have  been studied extensively, e.g. \cite{Every81,Shuvalov96}.  Shuvalov and Every 
\cite{Shuvalov97} provide a detailed study of the shape of the acoustic slowness surface of anisotropic solids near points of conical degeneracy, and show that 
there can be up to three pairs of lines of zero Gaussian curvature.  
It may be checked that 
the condition $\kappa=0$ is precisely Every's condition F, eq. (16) of \cite{Every81}.
The polarization vector fields near conical points display interesting properties \cite{Burridge67,alshits1979a,alshits1979b} and we refer the reader to the exhaustive analysis of Shuvalov \cite{Shuvalov98} for details.

\subsection{The wave surface}

Let the receiver position be defined in the new coordinate system as
\beq{0.07}
{\bf x} = r {\bf e}_\theta + z {\bf e}_3,  
\eeq
where ${\bf e}_\theta$ is orthogonal to ${\bf e}_3$ and defined by eq. \rf{502}
with $\phi$ replaced by $\theta$. 
The contribution to the  fundamental solution \rf{0.03} from the neighborhood
of the conical point  follows by approximating the function ${\bd F}$ of \rf{0.03b} 
locally   \cite{Burridge67}.  As mentioned above,  the gradient 
 $ \nabla v(\bd{\xi})$ equals the wave velocity ${\bd c}$, and therefore $ |\nabla v(\bd{\xi})| = c$, the magnitude.  The wave velocity satisfies  
  $\bd{\xi}\cdot {\bd c}=1$ \cite{Musgrave}, while near the conical point the approximations
  $\xi \approx \xi_0$ and  $\hat{\bd{\xi}}\cdot \hat {\bd c}\approx \cos \alpha $ hold. 
  The first is clear, while the second is a consequence of the fact that the slowness surface is to leading order conical.  Hence, $c \approx 1/(\xi_0 \cos \alpha )$. 
  The polarization has the form ${\bd A} = {\bd e}_{\pi/2 - \phi /2}$ (eq. (7.15) of \cite{Burridge67})  which together with the approximation for $ |\nabla v(\bd{\xi})| $ leads to  
\beq{1.01}
{\bf G} \approx -\, \frac{\xi_0\, \cos\alpha }{16\, \pi^2}   \,  \int_{S_c} \dd S  \, 
\begin{bmatrix} 
1-\cos \phi &\sin \phi &0\\
\sin \phi &1+\cos \phi &0\\
0&0&0
 \end{bmatrix} \,
\delta'(t- \bd{\xi} \cdot {\bf x} ).  
\eeq
Here $S_c$ denotes  the slowness surface in the vicinity of the conical point.  Also, the matrix elements are referred to the local basis 
$\{ {\bf e}_1,{\bf e}_2,{\bf e}_3\}$. 
It is important to note that $S_c$  is single-valued and that the two distinct quasi-transverse slownesses are included by virtue of the fact that both $\zeta<0$ and 
$\zeta >0$ must be considered in the integral.  This implies that the neighboring sheets at coincident values of  $|\zeta|$ correspond to ${\bd e}_{\phi}$ and ${\bd e}_{\phi +\pi}$ 
and the associated polarizations are (as above) 
${\bd A} = {\bd e}_{\phi /2}$ and 
${\bd A} = {\bd e}_{\phi /2 + \pi /2}$, respectively.   This formulation obviates the need to distinguish the two almost degenerate sheets.  
We note that the surface measure is  
$ \dd S = \big[ (1+ \gamma_\zeta^2)\gamma^2 
+\gamma_\phi^2\big]^{1/2}\, \dd \zeta\, \dd \phi  $,
where  $\gamma_\zeta \equiv \partial \gamma/\partial \zeta = \beta + \kappa (\phi)\zeta$ and $\gamma_\phi \equiv \partial \gamma/\partial \phi = \frac12 \kappa '(\phi)\zeta^2$.

The difficulty in evaluating \rf{1.01} arises from the 
phase term 
\beq{1.02} 
\bd{\xi} \cdot {\bf x}  = \xi_0 \, z + \zeta z + \gamma(\zeta,\phi)\, r
\cos(\phi-\theta) , 
\eeq
which follows from  \rf{0.07} and the local approximation \rf{0.06}.
As Table 1 demonstrates, the slowness surface curvature
parameter $\kappa $ defined in \rf{1.03} 
varies with $\phi$, and can change sign.

\subsubsection{Local approximation of the wave surface}

The wave (or group) velocity vector $\bd c$ is normal to the slowness surface and satisfies 
$\bd{c} \cdot \bd{\xi} = 1$ . 
These properties, combined with the slowness in eq. \rf{0.06} imply
\beq{503}
\bd{c} = \big( \gamma_\zeta {\bf e}_3 - {\bf e}_\phi +(\gamma_\phi /\gamma )
{\bf e}_{\phi + \frac{\pi}{2}}
\big)/ [ ({\xi}_0 +\zeta) \gamma_\zeta  - \gamma].
\eeq
In order to derive eq. \rf{503}, first note that the 
partial derivatives of the slowness follow from eqs. \rf{0.06} and \rf{502}    as 
$\bd{\xi}_\zeta  =  {\bf e}_3+ \gamma_\zeta {\bf e}_\phi $ and 
$\bd{\xi}_\phi  =  \gamma_\phi {\bf e}_\phi + \gamma 
{\bf e}_{\phi + \frac{\pi}{2}} $. These are tangent vectors to the slowness surface, and  the direction of $\bd{c}$ is therefore defined by their cross product.  This may be readily evaluated using the orthonormal properties of the triad 
$\{ {\bf e}_\phi , {\bf e}_{\phi + \frac{\pi}{2}}, {\bf e}_3 \}$. Then using eq. $\bd{c} \cdot \bd{\xi} = 1$ to fix the magnitude, we arrive at the expression for $\bd{c}$.

Consider a field point in the direction of the wave velocity: 
\beq{505}
\bd{x} \equiv z {\bf e}_3  + r \bd{e}_\perp  = t \bd{c}, 
\eeq
where $\bd{e}_\perp$ is the unit vector in the direction of  the projection of $\bd{c}$ on the plane perpendicular to the ${\bf e}_3$ direction (the [111] axis).  It follows from eq.  \rf{503} as 
$\bd{e}_\perp = \big( - {\bf e}_\phi +(\gamma_\phi /\gamma )
{\bf e}_{\phi + \frac{\pi}{2}}
\big)/ \sqrt{ 1 + (\gamma_\phi /\gamma )^2} $. 
Equating components in the  directions ${\bf e}_3  $ and $\bd{e}_\perp$ yields two relations, which we write as
\beq{507}
t = {\xi}_0 z + \zeta z - \frac{\gamma r}{\sqrt{ 1 + (\gamma_\phi /\gamma )^2} } ,
\qquad \quad 
\frac{z}{r}  = \frac{\gamma_\zeta}{\sqrt{ 1 + (\gamma_\phi /\gamma )^2} }.
\eeq

These identities are exact and provide implicit relations between the field point and the location of the point on the slowness surface, characterized by the variables
$\zeta $ and $\phi$ in the vicinity of the conical point. In particular, we have not 
yet invoked  the parabolic approximation \rf{1.03}.  Use of this allows us to to solve for the position on the slowness surface associated with a  field position lying near the circular boundary of the lid in physical space.  The edge of the lid is defined by 
\beq{510}
z-r\beta = 0, \qquad \mbox{where}\quad \beta = \cot \alpha . 
\eeq

%%%%%%%%%%%%%%%%%%%%%%%%%%%%%%%%%%%%%%%%%%%%%%%%
%\newpage 
\small
\begin{center}
Table 1.  The cone angle $\alpha$ and the nondimensionalized curvature 
${\bar \kappa}(\phi) \equiv \kappa(\phi)\, \xi_0/\beta $ follow from equations \rf{0.065} and
\rf{1.035} and $\beta = \cot\alpha$.  
Here     $\kappa_0 = \kappa(0)$,  $\kappa_1 =\kappa(\pi/3)$ if $\mu_2-\mu_1 >0$ with the definitions reversed if $\mu_2-\mu_1 <0$. 
Cubic elastic constants  from Musgrave 
 \cite{Musgrave}. % , except for $*$ \cite{Landolt}. 
\\
\vspace{.25in}
\begin{tabular}{|l|r|r|r|}%{|p{1.4in}|p{.56in}|p{.56in}|p{.56in}|p{.56in} |p{.56in}|p{.70in}|}
\hline\hline %&&& \\
\small Material &  $\alpha \, (deg)$ & ${\bar \kappa}_0 $ & $ \kappa_1  /\kappa_0$ 
\\  %&&& \\ 
\hline
%\small \\ 
diamond	&	5.32	&	122.03	&	1.33	 \\							
aluminum	&	5.41	&	117.93	&	1.36	\\							
silicon	&	12.61	&	22.14	&	1.70	\\							
germanium	&	14.39	&	16.94	&	1.77	\\							
silver chloride	&	14.82	&	10.61	&	-0.69	\\		
potassium chloride	&	20.46	&	4.29	&	-3.58	\\									
iron	&	25.02	&	5.11	&	2.35	\\							
nickel	&	25.38	&	4.93	&	2.37	\\							
gold	&	28.56	&	3.68	&	2.87	\\							
silver	&	29.59	&	3.35	&	2.79	\\							
copper	&	30.95	&	2.96	&	2.83	\\							
potassium	&	42.16	&	1.00	&	4.23	\\							
sodium 	&	44.84	&	0.72	&	4.34	\\							
$\beta$-brass	&	45.28	&	0.68	&	5.10	\\							
lithium	&	53.01	&	0.10	&	20.33	\\							
\hline
%\\ \hline
\end {tabular}
\end {center}
\vspace{.2in}
\normalsize
%%%%%%%%%%%%%%%%%%%%%%%%%%%%%%%%%%%%%%%%%%%%%%%%

The idea is to solve  eq.  $\rf{507}_2$ for $\zeta$ using \rf{1.03} with the assumption that 
$z-r\beta \approx 0$.  This amounts to finding solutions to leading order in the small parameter $\zeta$.  Thus, we find 
\beq{511}
\zeta = \frac{z-r\beta}{\kappa r} + \mbox{O}\big( (z-r\beta)^2\big) . 
\eeq
Substituting into \rf{507} and again expanding in $\zeta$ gives
\beq{512}
t=  {\xi}_0 z + \frac{(z-r\beta)^2}{2\kappa r} + \mbox{O}\big( (z-r\beta)^3\big). 
\eeq
The critical parameter in the previous two results is the curvature $\kappa (\phi )$, which is evaluated at the value of $\phi$ defined implicitly by eq. \rf{505}.   The expansion in terms of $(z-r\beta)$ fails if $\kappa$ vanishes, and clearly a different scaling is necessary near values of $\phi$ at which $\kappa (\phi )=0$.   We explicitly avoid such directions.  

For the remainder of the paper we adopt the major simplification that the curvature $\kappa$ is independent of $\phi$ and hence $\gamma$ is azimuthally symmetric, 
\beq{1.04} 
 \gamma =  \gamma(\zeta). 
\eeq
The removal of the variation in curvature with azimuth 
    leads to tractable expressions for the Green's function, ones that are more detailed than those of Burridge \cite{Burridge67}, yet which  retains the essential physics.  At the same time, the approximation is not overly excessive in the sense that  we expect  all directions of observation  will display  the same general properties found here.  The  exceptions are  for receiver positions defined by the  possible parabolic  lines on the slowness surface along which $\kappa (\phi )$ vanishes, which is outside the scope of the present analysis.

\section{Evaluation of the Green's function}\label{sec4}

\subsection{Reduction to a single integral}

As discussed  we restrict attention to the case of $\kappa$ independent of 
$\phi$. 
%and $\kappa >0$   %         ???????????????????????
Under this assumption the azimuthal integral in \rf{1.01} may be 
effected.  First change from $\phi$ to $\psi = \phi-\theta$, and noting
that $
\dd S = |\gamma(\zeta)|\, \sqrt{1+\gamma_\zeta^2}\, \dd \zeta\, \dd \psi $,
we obtain 
\begin{align}\label{1.06}
{\bf G} &\approx - \frac{  \xi_0 \cos\alpha }{8\pi}\,   \int \dd \zeta\,
|\gamma(\zeta)|\, \sqrt{1+\gamma_\zeta^2}\,  
\nonumber \\ & \quad \times 
\int \frac{\dd \psi}{2\pi}
\begin{bmatrix} 
1-\cos \theta\, \cos\psi  &\sin \theta\,\cos\psi &0\\
\sin \theta \,\cos\psi&1+\cos \theta\,\cos\psi &0\\
0&0&0
 \end{bmatrix} \,
\delta'(T- R\,\cos \psi ) .
\end{align}
Here, 
\beq{1.07} 
T=t -  \xi_0 \, z - \zeta z ,\qquad R=  \gamma(\zeta)\, r.
\eeq
The integral can be performed using the identity  
\beq{1.08} 
 \int_0^{2\pi} \frac{\dd \psi}{2\pi} \, f(\cos\psi)\,
\delta'(T- R\,\cos \psi )
= \left. \frac{1}{\pi R^2} \,
\frac{ \dd }{ \dd x}
\left(  \frac{f(x)}{\sqrt{1-x^2}}\right) \right|_{x=T/R}~
H(R^2-T^2),  
\eeq
where $H$ is the Heaviside step function ($H(t)=0,\frac12 , 1$ for $t<0,=0,>0$, respectively) with the result, 
\beq{1.09}%\begin{align}\label{1.09}
{\bf G} \approx - \frac { \xi_0\, \cos\alpha}{8\,\pi^2} \,   \int \dd \zeta\,
|\gamma(\zeta)|\, \sqrt{1+\gamma_\zeta^2}\,  
%\nonumber \\ & \quad 
\begin{bmatrix} 
T-R\, \cos \theta  &R\, \sin \theta &0\\
R\,\sin \theta&T+R\, \cos \theta &0\\
0&0&0
 \end{bmatrix}  \,
\frac{H(R^2-T^2)}{(R^2-T^2)^{3/2}} . 
\eeq% \end{align}

%Note that 
%\beq{1.10}R^2-T^2 = \left[ \frac{\kappa}{2}\,r\,\zeta^2 + (z+r\beta)\,\zeta -(t-\xi_0\,z)\right]\, 
%\left[ \frac{\kappa}{2}\,r\,\zeta^2 - (z-r\beta)\,\zeta +(t-\xi_0\,z)\right]. \eeq
Define two new dimensionless parameters
\beq{1.11}
\tau = \frac{2\kappa r}{(z+r\beta)^2}\, (t-\xi_0\,z) ,\qquad \qquad 
\eta = \frac{z-r\beta}{z+r\beta}. 
\eeq
$\tau$ is a reduced time variable relative to the time of arrival of the flat lid  wavefront, and $\eta$ defines distance from the lid edge, with $\eta >0$ $(<0)$ if the observer lies inside (outside) the lid, see Figure \ref{fig2}. 
Let $x = \kappa   r \zeta/(z+r\beta)$, then it may be checked that 
\beq{1.13}
R^2-T^2 = \frac{(z+r\beta)^4}{(2\kappa  r)^2} \,
\left[ (x+1)^2 - (1+\tau)\right]\,
 \left[ (x-\eta)^2 - (\eta^2-\tau)\right].
\eeq
There are four roots to $R^2 - T^2 = 0$, at 
$x = \eta \pm (\eta^2-\tau)^{1/2}$ and $x = -1 \pm (1+\tau)^{1/2}$. 
Points near the edge of the lid are characterized by 
$|\eta |\ll 1$, and the associated wavefronts from the lid and the 
``regular'' wave arrive at times such that the time parameter 
$\tau$ is also small.  Of the four roots, all are at values of 
$x$ that are asymptotically small except the single spurious
root at $x = -1 - (1+\tau)^{1/2} \approx -2$. The origin of this root lies 
with the parabolic approximation to the sheets of the slowness surface, and
it is not a real effect in elasticity.  We therefore ignore this root
and replace \rf{1.13}  by the cubic approximation 
\beq{1.15}
R^2-T^2 = \frac{(z+r\beta)^4}{(2\kappa r)^2} \,
g(x),
\qquad \mbox{with}\quad 
g(x) =  (2 x-\tau )\,
\left[ (x-\eta)^2 - (\eta^2-\tau)\right].
\eeq
We make the further approximation 
$ \sqrt{1+(\gamma_\zeta)^2}\approx 
\csc\alpha $, 
which is valid near the conical point.  Then noting that 
\begin{subequations}
\begin{align}\label{1.17}
\frac{\zeta\, T}{(R^2-T^2)^{3/2}} 
&= -\partial_z\, \frac{1}{(R^2-T^2)^{1/2}} , \\
\frac{\zeta\, R}{(R^2-T^2)^{3/2}} 
& =  -\frac{1}{r}\, \partial_\beta \, \frac{1}{(R^2-T^2)^{1/2}} 
%\nonumber \\ & 
\approx  -\frac{1}{\beta}\, \partial_r \, \frac{1}{(R^2-T^2)^{1/2}},
\end{align}
\end{subequations}
where $\partial_z $ is $\frac{\partial}{\partial_z} $, etc., 
we deduce from eqs. \rf{1.09} through \rf{1.15} that 
\beq{1.18}
{\bf G} \approx \frac{\beta\xi_0}{8\,\pi}   \,
 \begin{bmatrix} 
\beta\,\partial_z -  \cos \theta \,\partial_r &\sin \theta \,\partial_r &0\\
\sin \theta \,\partial_r &\beta\,\partial_z +  \cos \theta \,\partial_r &0\\
0&0&0
 \end{bmatrix}  \,
\frac{I(\tau , \eta) }{z+r\beta} \, ,
\eeq
where $I$ is the  integral 
\beq{1.19}
I (\tau , \eta)= \frac{2}{\pi} \, \int_{-\infty}^{\infty} \dd x\,
\sgn (x)\, \frac{H(g(x))}{\sqrt{g(x)}} . 
\eeq

\subsection{Evaluation of the integral}

The integral \rf{1.19} can always be reduced to 
a sum of elliptic integrals, which in turn  depend upon the number of 
roots of $g(x)=0$.  Because $g(x)$ is cubic it can 
have either one or three real roots.  
Two distinct cases must be distinguished:
\begin{subequations}
\begin{align}\label{1.20}
  \tau < \eta^2, & \qquad {\rm three~ roots}  \\
 \quad  \tau > \eta^2,& \qquad{\rm one~ root}.   \label{1.20b}
\end{align}
\end{subequations}
It is useful to introduce two new parameters, 
\beq{1.25}
\Delta\, \tau = \tau - \eta^2,
\eeq
which is the time relative to the curved ``regular'' wavefront, and 
\beq{1.26}
\tan \Phi = \frac{ \eta -\frac{ \tau}{2} }{\sqrt{|\Delta \tau}|},
\quad -\frac{\pi}{2} < \Phi < \frac{\pi}{2} . 
\eeq
%Note that $\eta$   is positive if the observer lies within the cone $r = z\cot \alpha $ and negative outside. 
We examine the two cases \rf{1.20} and \rf{1.20b} separately. 

\subsubsection{Before the regular front}
We  consider $\tau < \eta^2$, for which there are three roots to $g(x)=0$. 
Denote the roots by $a>b>c$, then it follows from \rf{1.15} 
that $abc = \tau^2 /2 >0$.  The roots are therefore either all positive,
or two are negative and one positive. In either case, 
\rf{1.19} becomes 
\beq{1.33}
I =  \frac{\sqrt{2}}{\pi} \, \left[ 
\sgn (c) \, \int_c^b 
\frac{\dd x }{\sqrt{ (x-a)(x-b)(x-c) }} 
+ \int_a^\infty 
\frac{\dd x }{\sqrt{ (x-a)(x-b)(x-c) }} \right]. 
\eeq
Equations (3.131) of \cite{Gradshteyn} imply that the two integrals in \rf{1.33} 
are identical in value and hence they cancel if $c <0$.  Simplifying the 
integrals gives 
\beq{1.34}
I =  \frac{4\sqrt{2}}{\pi} \, 
\frac{H(c)}{\sqrt{ a-c }} \, 
K\left( \sqrt{ \frac{b-c}{a-c} } \right),
\eeq
and $K$ is the complete elliptic integral of the first kind,
\beq{1.23}
K( k) =   \int_0^\frac{\pi}{2}
\frac{ \dd u }{\sqrt{1 - k^2\, \sin^2 u}}. 
\eeq
Referring to \rf{1.15} we see that $c >0$ iff both $\tau $ and $\eta$ are 
positive.  Hence, replacing $a$, $b$ and $c$ by the appropriate roots
$\eta + \sqrt{\eta^2 - \tau}$, $\eta - \sqrt{\eta^2 - \tau}$ and 
$ \frac{\tau}{2}$, respectively, yields
\beq{1.35}
I (\tau, \eta) =  %\frac{}{\pi} \, 
\frac{4\sqrt{2}\, H(\tau)\, H(\eta)}{ \pi
\sqrt{ \eta -\frac{\eta^2}{2}  -\frac{\Delta \tau}{2} 
+ \sqrt{ -\Delta \tau}  }} \, 
K\left( \sqrt{ \tan (\Phi -\frac{\pi}{4})} \right) ,
\eeq
where $\Delta \tau $ is defined in  \rf{1.25} and $\Phi$ in \rf{1.26}.

\subsubsection{After the regular front}
We now consider the situation when   \rf{1.20b} holds.  In this case  $g(x)=0$  has only one root at $x
=\tau /2$ and it follows, using equation (3.138.7) of \cite{Gradshteyn}, that 
\beq{1.21}
I=  \frac{\sqrt{2}}{\pi} \, \int_{\frac{\tau}{2}}^{\infty}
\frac{ \dd x}{\sqrt{(x-\frac{\tau}{2})[(x-\eta)^2 + (\tau-\eta^2)]}}
=   \frac{2}{\pi}
\,\sqrt{ \frac{2}{p}} \, K( q), 
\eeq
where
\beq{1.22}
p= \sqrt{(\eta - \frac{\tau}{2})^2 + (\tau-\eta^2)},\qquad 
q=\sqrt{ \frac{p+\eta- \frac{\tau}{2} }{2\, p}}.
\eeq
Substituting from \rf{1.25} and \rf{1.26}  reduces  \rf{1.21} to 
\beq{1.27}
I(\tau, \eta) =   \frac{2\sqrt{2}}{\pi}
\, \left[ \Delta \tau\,+\,  \left(\frac{\Delta \tau}{2}
-\eta + \frac{\eta^2}{2} \right)^2 \right]^{-1/4}\, \, 
K\left(  \cos (\frac{\Phi}{2} - \frac{\pi}{4})\right).
\eeq

\subsubsection{Summary of the two cases}

Combining \rf{1.35} and \rf{1.27} and using \rf{1.26} provides us with a concise
formula for $I(\tau, \eta)$:
\beq{1.37}
I =  \frac{2\sqrt{2}}{\pi} \, \frac{\sqrt{\cos\Phi}}{|\Delta\tau|^{1/4}}  \times 
\left\{ 
 \begin{array}{cr}
\sqrt{ 2 \sqrt{2} \sec (\Phi -\frac{\pi}{4}) } \, 
K\left( \sqrt{ \tan (\Phi -\frac{\pi}{4})} \right)\,\, H(\tau)\, H(\eta),
&\Delta\tau <0, \\ \\
K\left(  \cos (\frac{\Phi}{2} - \frac{\pi}{4})\right),
&\Delta\tau >0.
\end{array} \right.
\eeq
We are now in a position to apply these formulae and summarize the behavior of the Green's function. 

\section{Discussion of the wavefront singularities}\label{sec5}

The contribution to the Green's function from the neighborhood of the conical point is evidently contained in the functional form of eq. \rf{1.37}.  In addition to the flat lid response this should include the regular wavefronts associated with the curved surfaces of the two sheets near their point of intersection. 

\subsection{The flat lid wavefront}
The first thing we note is the step discontinuity in $I$ at $\tau = 0$
for $\eta >0$, i.e., inside the lid, see Figure \ref{fig2}. Using \rf{1.37} for $\Delta \tau <0$ and noting that $\Phi = \pi/4$ in this limit, implies 
\beq{1.38}
I(0+ , \eta) - I(0-, \eta) = \frac{2}{\sqrt{\eta}} . 
\eeq
Combined with \rf{1.11} and \rf{1.18} this gives the contribution to the 
Green's function from the flat lid,
\beq{1.39}
{\bf G} \approx \frac{\xi_0\, \beta^2 }{4\,\pi} \, 
\frac{H(t-\xi_0 z)}{\left( z^2-r^2\,\beta^2\right)^{3/2}} \, 
 \begin{bmatrix}
-z\, -r \beta \, \cos \theta  &r\beta \, \sin \theta  &0\\
r\beta \, \sin \theta  &-z +  r\beta \, \cos \theta  &0\\
0&0&0
 \end{bmatrix} , ~\eta>0.
\eeq
This  agrees with equation (7.22) of  \cite{Burridge67}.

\subsection{The curved  wavefronts near the lid edge}
We next consider the singularities at $\Delta\tau = 0$ for $\eta<0$
and $\eta>0$, but such that $|\eta|\ll 1$.   The latter implies that the 
field point lies close to the direction $z = r\, \cot\alpha$, the edge of the lid. 

For small $\eta$, we have 
\beq{1.40}
\partial_z \approx \frac{1}{2z}\, \partial_\eta, \qquad 
\partial_r \approx  - \, \frac{\beta}{2z}\, \partial_\eta,
\eeq
and therefore from \rf{1.18}, 
\beq{1.41}
{\bf G} \approx \frac{1}{16\,\pi} \, \frac{\xi_0 }{r^2} \,
{\bf A}\, {\bf A}^T\, \frac{\partial I}{\partial \eta},
\eeq
where the unit vector ${\bf A}$ is now ${\bf e}_{-\theta/2}$, 
\beq{1.42}
{\bf A}=
\cos \frac {\theta}{2} {\bf e}_1 
-\sin \frac{\theta}{2} {\bf e}_2.
\eeq
The behavior of $I$ as $\Delta \tau \rightarrow 0\pm$ depends upon how
$\Phi$ of eq. \rf{1.26} behaves in this limit.  We note that 
\beq{1.28}
\lim_{\Delta \tau \rightarrow 0\pm} \, \Phi = 
\frac{\pi}{2}\, \sgn \, \eta . 
\eeq

The two cases of the observer lying just outside or just inside the lid are considered in sequence. 

\subsubsection{The convex  wavefront outside the lid}
First, if $\eta<0$, it follows from \rf{1.27}, \rf{1.37} and \rf{1.28} that 
\beq{1.44}
I \approx \sqrt{ \frac{2}{|\eta|}} \, H(\Delta\tau), \qquad 
\frac{\partial I}{\partial \eta}
\approx 2\sqrt{ 2 \,|\eta|} \, \delta (\tau -\eta^2). 
\eeq
The behavior of the Green's function near to but outside of the lid edge  
is therefore, using \rf{1.11}, \rf{1.41} and \rf{1.44},
\beq{1.45}
{\bf G} \approx \frac{\xi_0 \beta^2}{4\pi} \, 
{\bf A}\, {\bf A}^T\, \, \frac{\sqrt{2\,|\eta|}}{\kappa r} \,
\delta \left(t-\xi_0z - \frac{(z-r\beta)^2}{2 \kappa r}\right),\quad \eta<0. 
\eeq
The delta function singularity is what one would expect from the 
locally convex shape of the slowness surface  and the fact that $\kappa >0$ is assumed, as occurs, for instance,  for iron in Figure \ref{fig0}. 
Equation \rf{1.45} also agrees with the general result of   eq.
(6.8) of \cite{Burridge67}.  It is interesting to 
note that the strength of the singularity decreases  as $\sqrt{|\eta|}$ as the field point approaches the lid edge. 

\subsubsection{The saddle-shaped  wavefront inside the lid}
We next consider the singularity at 
 $\Delta\tau =0 $ for $\eta>0$, for which we use  
\beq{1.29}
K(q) \approx \ln \frac{1}{\sqrt{1-q^2}}, \quad q\rightarrow 1.
\eeq
Equations \rf{1.35}  and \rf{1.28}  
together imply  that  for fixed but asymptotically small and positive 
$\eta$ and $\Delta \tau = {\rm o}(\eta)$, 
\beq{1.46}
I \approx   \frac{2}{\pi}\, 
\frac{1}{ \sqrt{ 2\, {\eta}}}\,\, 
 \ln \frac{1}{|\Delta \tau|} ,
\qquad \Delta \tau \rightarrow 0\pm. 
\eeq
Hence, 
\beq{1.47}
\frac{\partial I}{\partial \eta}
\approx  -\,   \frac{2}{\pi}\, 
\frac{ \sqrt{ 2 \,\eta}}{\Delta \tau +i0} ,
\eeq
and the Green's function is 
\beq{1.48}
{\bf G} \approx \frac{\xi_0 \beta^2}{4\pi} \, 
{\bf A}\, {\bf A}^T\,  \frac{\sqrt{2\,\eta}}{\kappa r} \,
\frac{1}{\pi \left(t-\xi_0z - \frac{(z-r\beta)^2}{2\kappa r} + i0
\right)}
,\quad  \eta>0 . 
\eeq
Again, this has the expected form for a singularity associated 
with a part of the slowness surface which is saddle shaped,
in agreement with equation (6.9) of \cite{Burridge67}. We note that the magnitude of the singularity also depends upon 
$\sqrt{\eta}$.

The difference between this case and the prior result eq. \rf{1.45}  is that the the normal to the slowness surface in the latter case is exterior to the cone, while for the present situation it lies inside the conical surface.  With reference to the picture for iron in Figure \ref{fig0}, the convex wavefront arises from the lower convex part, while eq. \rf{1.48} is associated with the 
interior of the conical "bowl".

\subsubsection{Summary of the response inside and outside the lid}

The three singularities associated with the lid wavefront and the 
wavefronts from the convex and saddle-shaped regions near the conical point are
described by equations \rf{1.39}, \rf{1.45} and \rf{1.48}, respectively. 
In summary, from \rf{1.45} and \rf{1.48}:
\beq{18}
{\bf G} \approx \frac{\xi_0 \beta^2}{4\pi} \, 
{\bf A}\, {\bf A}^T\, \, \frac{\sqrt{2\,|\eta|}}{\kappa r} \, 
\times 
\left\{
\begin{matrix}
\delta \left(t-\xi_0z - \frac{(z-r\beta)^2}{2\kappa r}\right),& \eta<0,
\\
& 
\\
\pi^{-1} \left(t-\xi_0z - \frac{(z-r\beta)^2}{2\kappa r} + i0
\right)^{-1}
,& \eta>0. 
\end{matrix}
\right.
\eeq

These results are, so far, predictable from the analysis in \cite{Burridge67}, combined with the parabolic form of the slowness surface   assumed here.  In that sense we have not yet derived any new forms of the Green's function, other than to specialize them to the parabolic local approximation of the slowness surface.  The remainder of the paper generates expressions that are not special cases of prior results. We conclude this section by considering the field right on the lid edge.  The transition from one regime to another is encapsulated in the general expressions of equations \rf{1.37} and \rf{1.41} and is examined next. In  section \ref{sec6} we derive uniformly asymptotic expressions that are valid across the lid edge.  

\subsection{The singularity in the direction of the lid edge}

The lid edge direction  $z=r\, \cot\alpha$ is particularly interesting, and the singularity there corresponds to the origin of the nondimensional  coordinates: 
$\eta =\tau = 0$. 
The general expressions in \rf{1.37} may be used.  In this limit the 
domain of the term for $\Delta \tau$ vanishes, implying zero field for
$\tau <0$.  The field after $\tau =0$  follows from \rf{1.37} and 
\rf{1.41}.  Using \rf{1.26}  it is straightforward to show that 
\beq{1.49}
\left.  \frac{\partial I}{\partial \eta}\right|_{\eta=0}
=    \frac{H(\tau)}{\pi \left(\tau\,+\, \frac{\tau^2}{4}\right)^{1/4}} \,
\left[ \frac{1}{\sqrt{\tau}}  \, 
\frac{\dd  K}{\dd  k} \left(\frac{1}{\sqrt{2}}\right) 
\,+\,  \frac{1}{\sqrt{2} \left(1\,+\, \frac{\tau}{4}\right)}  \, 
K \left(\frac{1}{\sqrt{2}}\right) \right]. 
\eeq
The derivative  of the elliptic integral in \rf{1.49} simplifies to 
$\pi/\sqrt{2}K$ for the particular argument 
$1/\sqrt{2}$ (see eqs. (8.123.2) and (8.122) in \cite{Gradshteyn}). 
Then using the identity $K(1/\sqrt{2}) = \Gamma^2(1/4)/4\sqrt{\pi}$ from 
(8.129.1) of \cite{Gradshteyn}, where $\Gamma(1/4) \approx 3.62561$, 
and making the replacement 
$1+\tau/4 \rightarrow 1$ in \rf{1.49}, we obtain
\beq{1.50}
\left. \frac{\partial I}{\partial \eta}\right|_{\eta=0}
\approx    \frac{H(\tau)}{\tau^{3/4}}\,\, 
\frac{2 \sqrt{2\pi}}{\Gamma^2 (1/4)}\,
\left[ 1 + \frac{\Gamma^4 (1/4)}{16\pi^2}\,
\sqrt{\tau} \right]. 
\eeq
The Green's function is therefore, from \rf{1.41},
\beq{1.51}
{\bf G} \approx \frac{\xi_0 \beta^2}{4\pi} \, 
{\bf A}\, {\bf A}^T\, \, \sqrt{\frac{\pi}{\beta}}\, \frac{(2\kappa )^{1/4} }{\Gamma^2 (1/4)}\,
  \frac{H(t-\xi_0 z)}{\kappa r [r(t-\xi_0 z)^3]^{1/4}} \,
\left[ 1 + \frac{\Gamma^4 (1/4)}{16\pi^2}\,
\bigg( \frac{ \kappa (t-\xi_0 z)}{2\beta z }\bigg)^{1/2} %\right\}^{\frac{1}{2}}\, 
\right],~~~\eta=0. 
\eeq

\section{A uniform solution in rescaled variables}\label{sec6}

The various singularities  and their interaction can be understood 
by  a rescaling suitable for the neighborhood of the lid edge.  Let 
\beq{2.01}
\tau = s\, \eta^2,
\eeq
where the time-like parameter is 
\beq{2.05}
s = 2\kappa r\, \frac{(t-\xi_0\,z)}{(z-r\beta)^2}   .
\eeq
We recast the results in terms of $s$ and $\eta$, where the region of interest is near the lid edge, i.e., $|\eta|\ll 1$ but  $s={\rm O}(1)$.  
The three wavefronts discussed in the previous section, the flat lid wavefront, and the ``regular" wavefronts inside and outside the cone $z-\beta r=0$ correspond to $(s=0, \eta >0)$, $(s=1, \eta >0)$ and $(s=1, \eta < 0)$, respectively. The rescaled time allows us to separate the three, and to observe their coalescence.  

Since $|\eta|\ll 1$,  \rf{1.26} implies that 
$\tan\Phi \approx \sgn\, \eta / \sqrt{|s-1|}$, and after some
simplification, \rf{1.37} reduces to 
\beq{2.015}
I (\tau, \eta)\approx    \frac{2}{\sqrt{|\eta|}}\, { J}(s,\sgn \eta),
\eeq
where 
\beq{2.02}
{ J}(s,\sgn \eta)=   \frac{\sqrt{2}}{\pi} \, \times
\left\{ 
 \begin{array}{lr}
0,&s<0,\\ \\
2\left( 1 + \sqrt{1-s}  \right)^{-1/2} \, 
K\left(  \frac{\sqrt{s}}{1 + \sqrt{1-s}} \right)\,H(\sgn\eta)
,&0<s<1,\\ \\
s^{-1/4} \, 
K\left(  \sqrt{ \frac{1}{2} + \frac{\sgn \,\eta}{2\,\sqrt{s}}}\right)
,& s > 1.
\end{array} \right.
\eeq
Note that $J$ only depends on $\eta$ via its sign. 
It follows that 
\beq{2.03}
\frac{\partial I}{\partial \eta}  \rightarrow \frac{-\sgn\,\eta}{|\eta|^{3/2}}
\left[ { J}  +4s\, \frac{\partial J}{\partial s}  \right], 
\qquad \mbox{for } \, \eta \ne 0, 
\eeq
and so the fundamental solution in the vicinity of the cone $z=r\cot\alpha$ (although not on it, i.e. $\eta \ne 0$) 
may be written, from \rf{1.41} and \rf{2.03}, as 
\beq{2.04}
{\bf G} \approx -\, \frac{\xi_0 \beta^2}{4\pi} \, 
{\bf A}\, {\bf A}^T\, 2z\,  \frac{ \sgn 
(z-r\beta)}{|z^2-r^2\beta^2|^{3/2}}\, 
\left[ { J}(s,\eta) +4s\, \frac{\partial J}{\partial s}(s,\eta) \right].
\eeq
%%%%%%%%%%%%%%%%%%%%%%%%%%%%%%%%%%%%%%%%%%%%%%%%%% Figure
\begin{figure}[htbp]
				\begin{center}	
				\includegraphics[width=3.6in , height=2.4in 					]{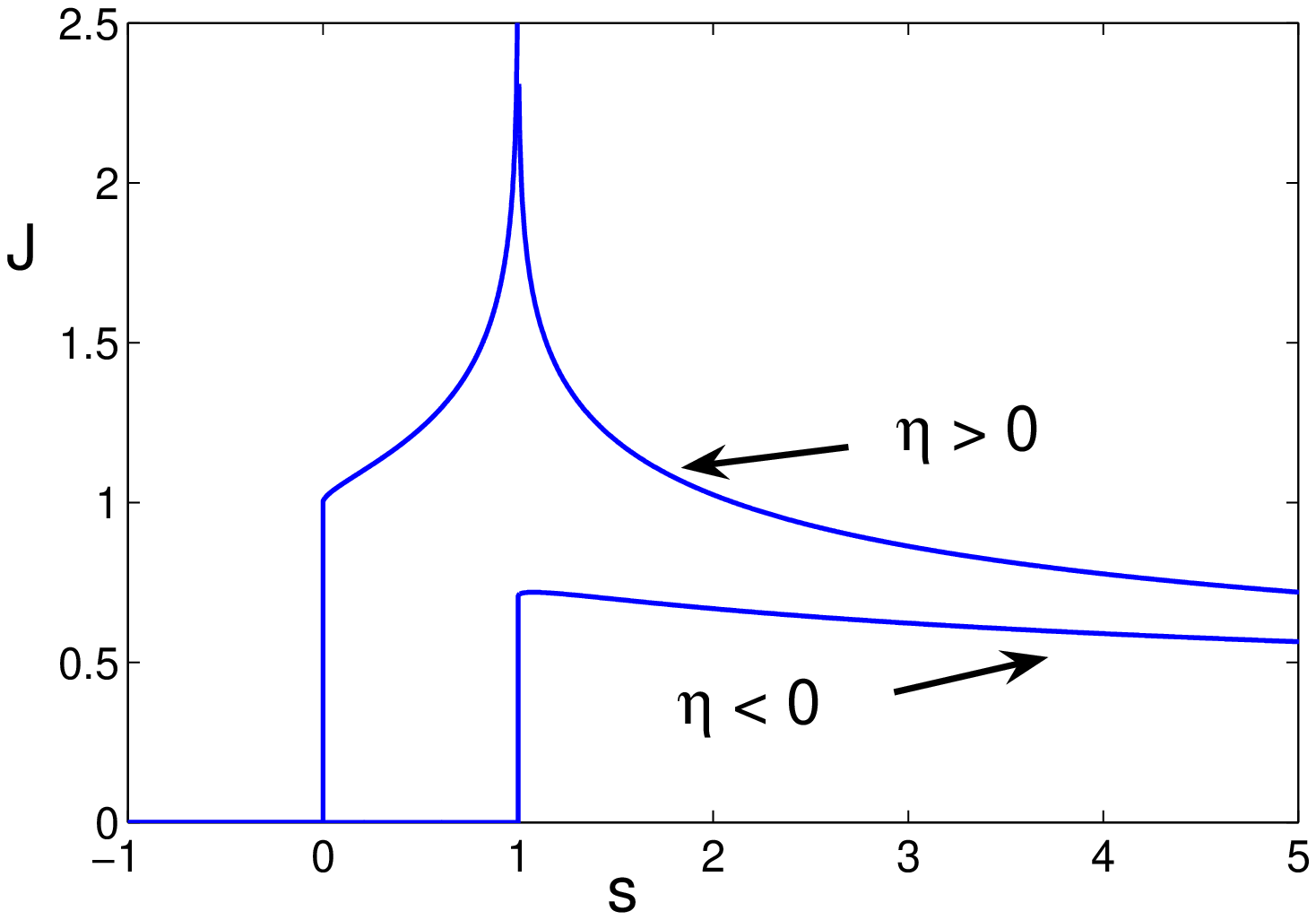} 
	\caption{The function $J(s,\sgn \eta )$ of eq. \rf{2.02}}
		\label{fig3} \end{center}  
	\end{figure}
%%%%%%%%%%%%%%%%%%%%%%%%%%%%%%%%%%%%%%%%%%%%%%%%%% 

The function $J$ is plotted in Figure \ref{fig3}.  Note the step 
singularity at $s=0$ for $\eta>0$, corresponding to the 
lid wavefront of \rf{1.39}.  The singularities at $s=1$ are 
a step for $\eta<0$, which results in the  delta pulse 
of \rf{1.45}, and a logarithmic singularity 
for $\eta>0$, corresponding to the $1/t$ wave in \rf{1.48}.

\begin{figure}[htbp]
				\begin{center}	
				\includegraphics[width=3.6in , height=2.4in 					]{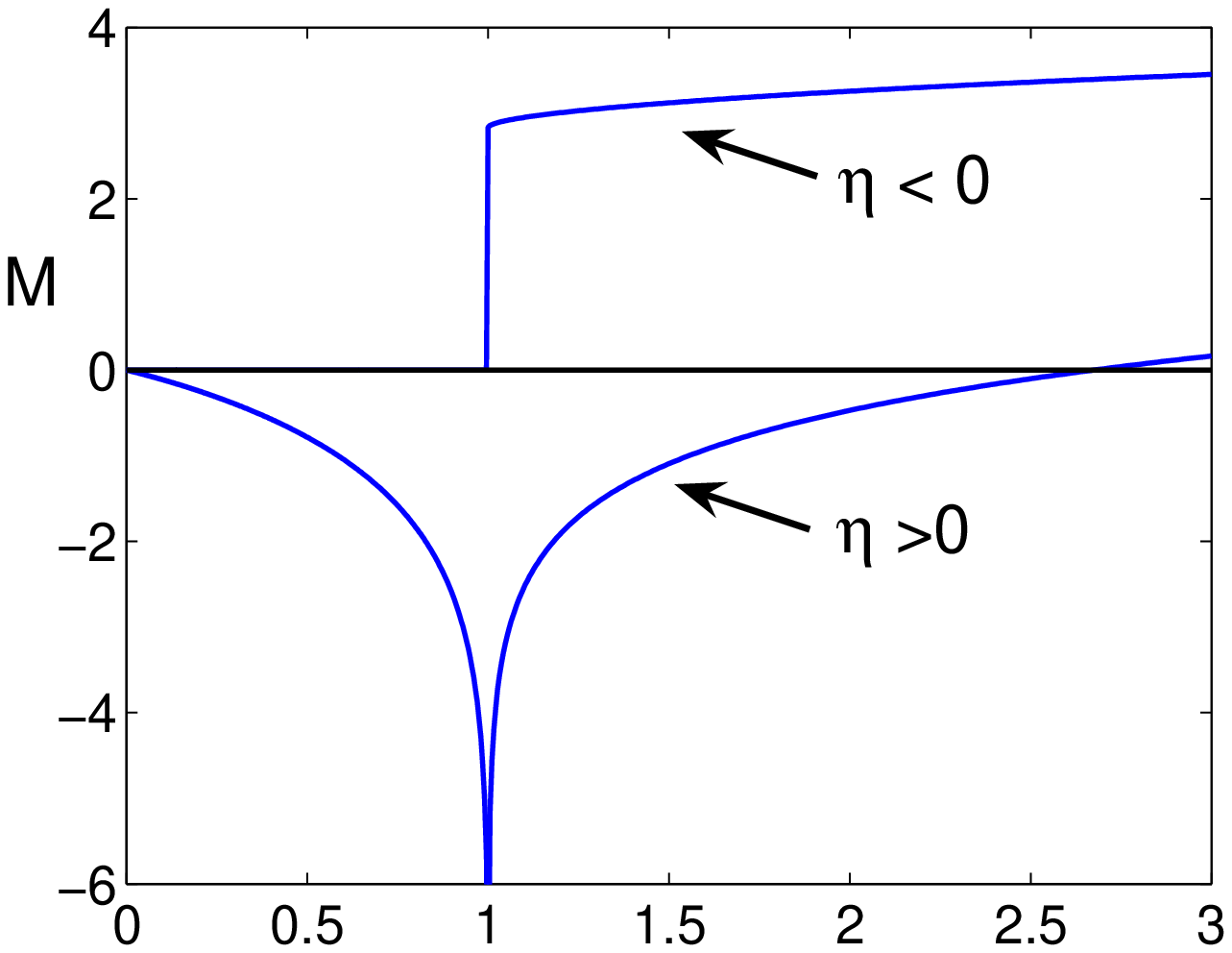} 
	\caption{The function $M(s,\sgn \eta )$ of eq. \rf{2.06}}
		\label{fig4} \end{center}  
	\end{figure}

The propagating singularities are all contained in the 
expression $(J+4s\partial J/\partial s)$ in \rf{2.04}.  Because this can exhibit 
delta function behavior which is difficult to display
on a plot, we look instead at the response from a step load,
{\it i.e.}, the solution to \rf{0.01} with $\delta(t)$ replaced
with $H(t)$.  Integrating \rf{2.04} gives the step response
near the edge of the lid as 
\beq{2.045}
{\bf G} \approx  \frac{\xi_0 \beta^2}{4\pi} \, 
{\bf A}\, {\bf A}^T\,  \frac{ \sqrt{|\eta|}}{2\kappa r} \, 
M(s,\sgn \eta) , 
\eeq
where 
\begin{align}\label{2.06}
M(s,\sgn \eta) &= -\sgn\, \eta\, \int_0^s
\left(  J +4s\, \frac{\partial J}{\partial s} \right)\, \dd s
\nonumber \\
&= \sgn\, \eta \, \left(  3\,  \int_0^s\, J\, \dd s -4\,s\,J \right).
\end{align}
The function $M$ is plotted in Figure \ref{fig4}.  For observers outside the cone, 
i.e. $z-r\beta <0$,  there is no signal until the arrival of the regular wavefront at $s=1$ $(\tau = \eta^2)$, and is  the step function instead of the delta response in 
eq. \rf{18}.   Inside the cone $(z-r\beta >0)$ the first, weak, signal is from the lid edge at $s=0$ $(\tau = 0)$ which has the form of a ramp function, as compared with the step in eq. \rf{1.39}.  This is followed by the regular wavefront at $s=1$ $(\tau = \eta^2)$ with a logarithmic singularity.  

%%%%%%%%%%%%%%%%%%%%%%%%%%%%%%%%%%%%%%%%%%%%%%%%%% Figure
%%%%%%%%%%%%%%%%%%%%%%%%%%%%%%%%%%%%%%%%%%%%%%%%%% 

The one remaining feature not evident in the uniform solution in terms of the $J$ and $M$ functions is the behavior right on the lid edge.   This follows by taking the limit $s\rightarrow \infty$ with $\eta \ne 0$.    Using \rf{2.02} and integrating by parts, gives \begin{align}\label{2.07}
M(s,\sgn\eta) - M(1,\sgn\eta)
 &= \frac{1}{\pi}\, \int_1^s\, 
\frac{\dd s}{s^{3/4}}\, 
\frac{1}{\sqrt{ 1 + \frac{\sgn \,\eta}{\sqrt{s}}}}
\, \frac{\dd  K}{\dd  k} \left(
\sqrt{ \frac{1}{2} + \frac{\sgn \,\eta}{2\,\sqrt{s}}} \right) 
\nonumber \\
&= \frac{8\,\sqrt{2\pi}}{\Gamma^2(1/4)}\, s^{1/4} \, +\,{\rm O}(1),
\quad s\rightarrow \infty . 
\end{align}
Thus  $M(s,\sgn\eta)$  becomes independent of $\sgn\,\eta$ for large $s$, 
and grows as $1.5255\, s^{1/4}$, which is consistent with the 
$1/t^{3/4}$ singularity along the directions
for which $\eta = 0$, see equation \rf{1.51}. 

\section{Conclusion}

We have revisited the problem of determining the waves emanating from the vicinity of the conical point in  materials with cubic symmetry.    The analysis is based on the general  formulation of Burridge \cite{Burridge67} and employs a higher order approximation to the slowness surface.   The key ingredient is the parabolic approximation  \rf{1.03}  for the vicinity of the conical point, which depends upon the  curvature $\kappa$ of eq. \rf{1.035}.   By assuming the curvature is constant in the azimuth the evaluation of the Green's function is reduced to a single integral \rf{1.19}.  This reproduces the flat lid singularity \rf{1.39} derived by Burridge \cite{Burridge67} as well as the regular wavefronts originating  near the conical point.   
The wavefront singularity on the lid edge, eq. \rf{1.51}, has been evaluated.  A uniformly valid representation is presented for the transition as the observation direction crosses the lid edge.  This involves a single function, based on rescaled space and time coordinates, that  displays the regular wavefront singularities and the flat lid singularity.  

\section*{Acknowledgment}
Helpful suggestions from A. L. Shuvalov are appreciated. 

%%%%%%%%%%%%%%%%%%%    bibliography   %%%%%%%%%%%%%%%%%%%%%%%%%%%%%%%%%%%%%
%			\bibliography{thermoelastic}

\providecommand{\bysame}{\leavevmode\hbox to3em{\hrulefill}\thinspace}
\providecommand{\MR}{\relax\ifhmode\unskip\space\fi MR }
\providecommand{\MRhref}[2]{%
  \href{http://www.ams.org/mathscinet-getitem?mr=#1}{#2}
}
\providecommand{\href}[2]{#2}

\end{document}